\documentclass[prd,showpacs,preprintnumbers,floats,floatfix,nofootinbib]{revtex4}

\usepackage{graphicx}
\usepackage{dcolumn}

\begin{document}

\title{The Efficiency of Gravitational Bremsstrahlung Production in the
Collision of Two Schwarzschild Black Holes}

\author{R. F. Aranha}
\email{rfaranha@cbpf.br}
\affiliation{Centro Brasileiro de Pesquisas F\'\i sicas-CFC\\ Rua Dr. Xavier
Sigaud 150, Urca, Rio de Janeiro CEP 22290-180-RJ, Brazil}

\author{H. P. de Oliveira}
\email{oliveira@dft.if.uerj.br}
\affiliation{{\it Instituto de F\'{\i}sica - Universidade do Estado do Rio de Janeiro }\\
{\it CEP 20550-013 Rio de Janeiro, RJ, Brazil}}

\author{I. Dami\~ao Soares}
\email{ivano@cbpf.br}
\affiliation{Centro Brasileiro de Pesquisas F\'\i sicas\\ Rua Dr. Xavier
Sigaud 150, Urca, Rio de Janeiro CEP 22290-180-RJ, Brazil}

\author{E. V. Tonini}
\email{tonini@cefetes.br} \affiliation{Centro Federal de Educa\c
c\~ao Tecnol\'ogica do Esp\'\i rito Santo\\ Avenida Vit\'oria,
1729, Jucutuquara, Vit\'oria CEP 29040-780-ES, Brazil}

\date{\today}

\begin{abstract}
We examine the efficiency of gravitational bremsstrahlung production
in the process of head-on collision of two boosted Schwarzschild
black holes. We constructed initial data for the characteristic
initial value problem in Robinson-Trautman spacetimes, that
represent two instantaneously stationary Schwarzschild black holes
in motion towards each other with the same velocity.
The Robinson-Trautman equation was integrated for these initial data
using a numerical code based on the Galerkin method. The final
resulting configuration is a boosted black hole with Bondi mass greater than the sum of the individual mass of each initial black hole. Two relevant aspects of the process are
presented. The first relates the efficiency $\Delta$ of the energy
extraction by gravitational wave emission to the mass of the final
black hole. This relation is fitted by a distribution function of
non-extensive thermostatistics with entropic parameter $q \simeq
1/2$; the result extends and validates analysis based on the
linearized theory of gravitational wave emission. The second is a
typical bremsstrahlung angular pattern in the early period of
emission at the wave zone, a consequence of the deceleration of the
black  holes as they coalesce; this pattern evolves to a quadrupole
form for later times.
\end{abstract}

\maketitle

\vspace{1cm}

\section{Introduction}

The collision of black holes, the collapse of stellar objects in the
process of formation of black holes as well as the evolution of
distorted black holes figure as promising sources of gravitational
waves. The importance of these issues lies in the fact that the
knowledge of gravitational waveforms originating from the above
processes will be of crucial importance for the recent efforts to
detect gravitational waves. However to date these issues still
remain far from completely understood and for most situations we
are forced to rely on approximation methods and numerical techniques
to obtain information on wave form patterns and radiative transfer
processes in the dynamics of gravitational wave
emission\cite{baiotti}.

An important astrophysical situation in which gravitational radiation is produced is that of a merger of two black holes\cite{pretorius}, in particular in a head-on collision, both with the same initial velocity. If the masses are distinct we should expect that the remnant of the collision would be a black hole with smaller velocity and the gravitational radiation produced to be typically bremsstrahlung in close analogy with electromagnetic bremsstrahlung.

Head-on collisions of two black holes were discussed in detail, both
numerically and semi-analytically, by Aninos et al. \cite{aninos},
Price and Pullin\cite{pp} and Gleiser et al.\cite{gleiser}. The
common feature of these papers is the use of Misner initial
data\cite{misner} to describe the initial configuration of two equal
mass nonspinning black holes. Their approach bridges numerical
relativity and perturbative techniques\cite{abrahams}~\cite{aninos1}
to extract the gravitational wave forms at the wave zone. In Ref.
\cite{aninos}, $l=2$ and $l=4$ waveforms at several radii of
extraction were exhibited for one initial data set, corresponding to
a proper distance between the throats for which no initial common
apparent horizon is present. Also, using the extracted wave forms,
the total gravitational wave energy output was calculated for six
initial data sets corresponding to six distinct values of the proper
distance between the throats. Initial data with small proper
distance between the throats were also considered (initially close
black holes) corresponding to a global initial apparent horizon. The
latter situation of close black holes was examined in Ref.
\cite{pp}, in which the presence of a global apparent horizon
allowed for the use of black hole perturbation theory; the computed
gravitational radiation was in accordance with the results of
numerical computations in \cite{aninos}. A comparison of the
different approaches in treating the head-on collision of two black
holes using Misner data was given in \cite {aninos2}. We should also
refer to Nicasio et al.\cite{nicasio} as a relevant reference in the
problem of head-on collision of two initially boosted black holes
using Brill-Lindquist type data\cite{brill}.
In this vein we will consider here the interaction of two boosted
black holes moving straight toward each other along the symmetry
axis with the same velocity, and modeled in the context of
Robinson-Trautman (RT) spacetimes\cite{rt}. In general the outcome is a
boosted black hole with smaller velocity than the initial velocity
of the two holes, and a larger amount of
mass-energy than the sum of the individual mass-energies of the
individual holes. In this process gravitational waves are emitted
while the black holes decelerate and coalesce, producing a pattern
typical of bremsstrahlung. RT dynamics is in the realm of characteristic
initial data evolution and Bondi's mass is adopted as
the total mass-energy definition.

The plan of the paper is the following.
In Section 2 the basic aspects of Robinson-Trautman spacetimes are
presented. Section 3 is devoted to a new construction of
Brill-Lindquist type initial data for Robinson-Trautman spacetimes
which can be interpreted as two initially boosted Schwarzschild
black holes; the initial data adapt properly to the initial value
problem on null cones as is the case in RT dynamics. The main
consequences of the dynamical evolution of these initial data are
discussed in Section 4, where the numerical integration is performed
using a code based on the Galerkin projection method. In Section 5
we summarize our main results and discuss their relevance and
limitations as compared to previous calculations of the problem in
the recent literature; an outline of future perspectives in this
subject is also done. Throughout the paper we use units
such that $8 \pi G=c=1$.

\section{Robinson-Trautman spacetimes}

Robinson-Trautman (RT) metrics are solutions of vacuum Einstein's
equations representing an isolated gravitational radiating system.
In a suitable coordinate system they have the form

{\small
\begin{eqnarray}
\label{eq1} ds^2&=&\Big({\lambda(u,\theta)- \frac{2~m_{0}}{r}+ 2 r
\frac{\dot{K}}{K}}\Big) d u^2+2du dr-r^{2}K^{2}(u,\theta)(d
\theta^{2}+\sin^{2}\theta d \varphi^{2}).
\end{eqnarray}}

\noindent Einstein's vacuum equations imply

{\small
\begin{equation}
\label{eq2}
\lambda(u,\theta)=\frac{1}{K^2}-\frac{K_{\theta \theta}}{K^3}+\frac{K_{\theta}^{2}}{K^4}-
\frac{K_{\theta}}{K^3}\cot \theta
\end{equation}}

\noindent and

{\small
\begin{equation}
\label{eq3}
-6 m_{0}\frac{\dot{K}}{K}+\frac{(\lambda_{\theta} \sin \theta)_{\theta}}{2 K^2 \sin \theta}=0.
\end{equation}}

\noindent In the above, a dot and a subscript $\theta$ denote
derivatives with respect to $u$ and $\theta$, respectively, and
$m_0$ corresponds to the Schwarzschild mass when $K=1$. Throughout
the paper we use units such that $8 \pi G=c=1$. The dynamics of the
gravitational field is totally contained in the function
$K(u,\theta)$ and governed by Eq. (\ref{eq3}), denoted RT equation.

The initial data problem for RT spacetimes belongs to the
class of characteristic initial value formulations as opposed
to the 1+3 formulation, according to the classification of
York\cite{york}. The degrees of freedom of the gravitational field
are contained in the conformal structure of parametrized 2-spheres
embedded  in a 3-spacelike hypersurface. For RT spacetimes
the function $K(u_{0},\theta)$ given in a characteristic surface
$u=u_{0}$ corresponds to the initial data to be evolved via 
the RT equation (\ref{eq3}).

In the semi-null local Lorentz frame given in \cite{oliv1} the
curvature tensor of RT spacetimes is expressed as

{\small
\begin{eqnarray}
\label{eq4}
R_{ABCD}=\frac{II_{ABCD}(u,\theta)}{r^3}+\frac{III_{ABCD}(u,\theta)}{r^2}+\frac{N_{ABCD}(u,\theta)}{r},
\end{eqnarray}}

\noindent where $II$, $III$ and $N$ are objects of Petrov-type $II$,
$III$ and $N$, respectively, displaying the peeling
property\cite{petrov}. Eq. \ref{eq4} shows that RT indeed is the
exterior gravitational field of a bounded configuration emitting
gravitational waves, and for large $r$ the spacetime looks like a
gravitational wave with propagation vector $\partial/\partial r$.
The curvature tensor components that contribute to the gravitational
degrees of freedom transverse to the direction of propagation of
the wave are

{\small
\begin{equation}
\label{eq5}
R_{0303}=-R_{0202}=-\frac{D(u,\theta)}{r}+{\cal{O}}\left(\frac{1}{r^2}\right),
\end{equation}}

\noindent where

{\small
\begin{eqnarray}
\label{eq6} D(u,\theta)&=&\frac{1}{2K^2} \partial_{u}
\left[\frac{K_{\theta \theta}}{K}-\frac{K_{\theta}}{K} \cot \theta-
2 \left(\frac{K_{\theta}}{K}\right)^2\right].
\end{eqnarray}}

\noindent The function $D$ contains all the information of the
angular, and time dependence of the gravitational wave amplitudes in
the wave zone.

Now the basic equations (\ref{eq2}), (\ref{eq3}) have the
stationary solution

{\small
\begin{equation}
\label{eq7} K(\theta)=\frac{K_0}{\cosh \gamma~+ \cos \theta~\sinh
\gamma}
\end{equation}}

\noindent where $K_0$ and $\gamma$ are constant, and as a
consequence $\lambda=K_0^{-2}$ (cf. Eq. \ref{eq2}). According to
Bondi and Sachs\cite{bondi} this solution can be interpreted as a
black hole boosted along the negative $z$-axis, with velocity
parameter $v= \tanh \gamma$ and mass function
$m(\theta)=m_{0}K^3(\theta)$. The total mass-energy of the
gravitational configuration results then in $M=m_{0}/2
\int^{\pi}_{0}K^3(\theta)\sin \theta d \theta =m_{0}\cosh
\gamma=m_{0}/\sqrt{1-v^2}$, where for the sake of convenience we
have set $K_0=1$. The parameter $\gamma$ is associated with the
rapidity parameter of a Lorentz boost given by the K-transformations
of the BMS group\cite{bondi}. The interpretation of (\ref{eq7}) as a
boosted black hole is relative to the asymptotic Lorentz frame which
is the asymptotic rest frame of the black hole when $\gamma=0$.

It will be important to exhibit a suitable expression for the total
mass-energy content of the RT spacetimes or simply the Bondi mass.
To accomplish such task it is necessary to perform a coordinate
transformation to a coordinate system in which the metric
coefficients satisfy the Bondi-Sachs boundary conditions (notice
that in the RT coordinate system the presence of the term
$2r\dot{K}/K$ does not fulfill the appropriate boundary conditions).
We basically generalize the procedure outlined by Foster and
Newman\cite{foster} to treat the linearized problem, whose details
can be found in Ref. \cite{oliv3,kramer}. The result of
interest is that Bondi's mass function can be written for any $u$ as
$M(u,\theta) = m_0 K^3(u,\theta) + \rm{corrections}$, where the
correction terms are proportional to the first and second Bondi-time
derivatives of the news function; furthermore at the initial null
surface $u=u_0$ these correction terms can be set to zero by
properly eliminating an arbitrary function of $\theta$ appearing in
the coordinate transformations from RT coordinates to Bondi's
coordinates. Thus the total Bondi mass of the system at $u=u_0$ is
given by

{\small
\begin{equation}
\label{eqBondi}
M(u_0)=\frac{1}{2}m_0\,\int_0^\pi\,K^3(u_0,\theta)\,\sin\theta d\theta.
\end{equation}
}

\section{Black Hole Initial Data for RT Spacetimes}
The object of this Section is to construct initial data $K(0,\theta)$
for the RT equation, representing two interacting black holes
instantaneously boosted. In order to accomplish this task
we will construct 3-dim initial data for two black holes
(that are similar to Brill-Lindquist(BL) data\cite{brill}) from
which we will extract the parametrized conformal factor $K(\theta)$ of
a family of two spheres embedded in this 3-dim geometry.
We will rely on the seminal paper by Misner\cite{misner} which will
also be a reference for our notation in the present Section.

Starting from bispherical coordinates\cite{arfken} in the 3-dim
Cartesian plane $\Sigma$, we are led to introduce the following
parametrization for Cartesian coordinates
\begin{eqnarray}
\nonumber
x&=&  \frac{a \sin \theta~ \sinh \eta}{\cosh \eta - \cos \theta~ \sinh{\eta}}~ \cos \varphi,\\
y&=&  \frac{a \sin \theta~ \sinh \eta}{\cosh \eta - \cos \theta~ \sinh{\eta}}~ \sin \varphi,~~~~~~~~~~ z>0\\
\nonumber
z&=&  \frac{a}{\cosh \eta - \cos \theta~ \sinh{\eta}},\\
\nonumber \label{eq3_1}
\end{eqnarray}
and
\begin{eqnarray}
\nonumber
x&=& - \frac{a \sin \theta~ \sinh \eta}{\cosh \eta + \cos \theta~ \sinh{\eta}}~ \cos \varphi,\\
y&=& - \frac{a \sin \theta~ \sinh \eta}{\cosh \eta + \cos \theta~ \sinh{\eta}}~ \sin \varphi,~~~~~~~~~~ z<0\\
\nonumber
z&=& - \frac{a}{\cosh \eta + \cos \theta~ \sinh{\eta}}.\\
\nonumber \label{eq3_2}
\end{eqnarray}
where $0 \leq \eta \leq \infty$, $0 \leq \theta \leq \pi$, $0 \leq \varphi \leq 2 \pi$.
In this parametrization, for each $\eta=\eta_0$ it corresponds two spheres, one 
at $z>0$ and the other at $z<0$. Also, for a given point $P$ of $\Sigma$ in $z>0$ fixed by
($\eta_{0},\theta_{0},\varphi_{0}$) there is an associated unique antipodal
point $P_a \rightarrow$ ($\eta_{0},\pi - \theta_{0},\varphi_{0}$) in the region $z<0$,
so that we go from one to the other by an inversion through the origin (cf. Fig. 1);
the Cartesian vector from the origin to the point $P:(x,y,z)$ has length
\begin{eqnarray}
r_{>}=a \sqrt{\frac{\cosh \eta + \cos \theta \sinh \eta}{\cosh \eta
- \cos \theta \sinh \eta}} \label{eq3_3}
\end{eqnarray}
while its corresponding antipodal point is a distance $r_{<}$ from the origin 
given by
\begin{eqnarray}
r_{<}=a \sqrt{\frac{\cosh \eta - \cos \theta \sinh \eta}{\cosh \eta
+ \cos \theta \sinh \eta}} \label{eq3_4}
\end{eqnarray}
\noindent The usefulness of this parametrization will become clear in
what follows. The surface $z=0$ corresponds to $\eta= \infty$. We
note that the Cartesian coordinates are continuous functions, with
continuous second derivatives, of ($\eta,\theta,\varphi$).
Singularities occurring are the usual singularities of a spherical
coordinate system.

\begin{figure}
\begin{center}
\rotatebox{270}{\includegraphics*[height=8cm,width=8cm]{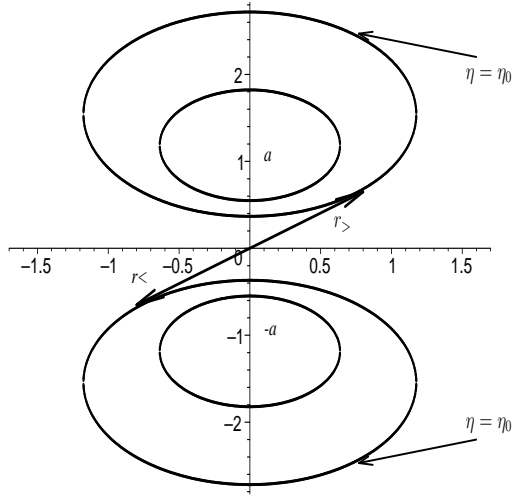}}
\caption{Projection of the spheres parametrized by $\eta_0$ into the plane $xz$.
The Cartesian vectors $\textbf{r}_{>}$ and $\textbf{r}_{<}$ localize an arbitrary point
of the spheres $\eta=\eta_0$ in $z>0$ and $z<0$, respectively.
}
\label{fig1}
\end{center}
\end{figure}

The flat space line element $ds^2=(dx)^2+(dy)^2+(dz)^2$ is expressed
in the above parametrization as

{\small
\begin{eqnarray}
ds^2=a^2~\frac{1}{(\cosh \eta \mp \cos \theta \sinh \eta)^2}\Big[d
\eta^2 + \sinh^2{\eta}(d\theta^2+\sin^2 \theta d \varphi^2) \Big]
\label{eq3_5}
\end{eqnarray}}

\noindent for $z>0$ and $z<0$, respectively.

We now take $\Sigma$ as a spacelike surface of initial data, with geometry
defined by the line element

{\small
\begin{eqnarray}
ds^2=K^2(\eta+\eta_0,\theta)~\Big[d \eta^2
+\sinh^2({\eta+\eta_0})(d\theta^2+\sin^2 \theta d \varphi^2) \Big].
\label{eq3_6}
\end{eqnarray}}

\noindent By assuming time-symmetric data (namely, $\Sigma$ a maximal slice)
we obtain that the Hamiltonian constraints reduce to
$^{(3)}R=0$. In this case the metric (\ref{eq3_6}) must satisfy the equation

{\small
\begin{eqnarray}
 -2 \frac{K_{\theta \theta}}{K^3} +\Big( \frac{K^{\prime
2}}{K^4}- 2 \frac{K^{\prime \prime}}{K^3} -4 \frac{K^{\prime }}{K^3}
\coth (\eta+\eta_0) \Big)~ \sinh^2 (\eta+\eta_0) & & \nonumber \\
- 2 \frac{K_{\theta}}{K^3} \cot \theta +\frac{K_{\theta}^{2}}{K^4}-3
\frac{\sinh^2 (\eta+\eta_0)}{K^2}=& &0, \label{eq3_7}
\end{eqnarray}}

\noindent where a prime denotes derivative with respect to $\eta$.
We note that the substitution $K=\Phi^2$ reduces (\ref{eq3_7}) to
a Laplace equation. Obviously the flat space solution

{\small
\begin{eqnarray}
K(\eta+\eta_0,\theta)=\frac{1}{\cosh (\eta+\eta_0) \mp \cos \theta
\sinh (\eta+\eta_0)} \label{eq3_8}
\end{eqnarray}}

\noindent satisfies Eq. (\ref{eq3_7}), from which it follows that
the function

{\small
\begin{eqnarray}
K(\eta,\theta)=\left(\frac{\alpha_1}{\sqrt{\cosh (\eta+\eta_0) -
\cos \theta \sinh (\eta+\eta_0)}} +\frac{\alpha_2}{\sqrt{\cosh
(\eta+\eta_0) + \cos \theta \sinh (\eta+\eta_0)}} \right)^2
\label{eq3_9}
\end{eqnarray}}

\noindent is a nonflat solution of (\ref{eq3_7}), where $\alpha_1$,
$\alpha_2$ and $\eta_0$ are arbitrary positive constants.
We may interpret the non-flat 3-dim metric defined
by (\ref{eq3_9}),

{\small
\begin{eqnarray}
ds^2= a^2~K^2(\eta,\theta)~\Big[d \eta^2
+\sinh^2{(\eta+\eta_0)}(d\theta^2+\sin^2 \theta d \varphi^2) \Big],
\label{eq3_10}
\end{eqnarray}}

\noindent as a BL-type solution given in bispherical coordinates.
In fact a straightforward manipulation shows that the metric
(\ref{eq3_10}) can be rewritten as

{\small
\begin{eqnarray}
\nonumber
ds^2=\frac{1}{2}\left(~{\alpha_2}+{\alpha_1}~ \sqrt{\frac{\cosh (\eta+\eta_0) +
\cos \theta \sinh (\eta+\eta_0)}{\cosh (\eta+\eta_0) - \cos \theta \sinh (\eta+\eta_0)}}~\right)^4~ds^{2}_{{\rm flat(+)}}\\
+\frac{1}{2}\left(~{\alpha_1}+{\alpha_2}~ \sqrt{\frac{\cosh
(\eta+\eta_0) - \cos \theta \sinh (\eta+\eta_0)}{\cosh (\eta+\eta_0)
+ \cos \theta \sinh (\eta+\eta_0)}}~\right)^4~ds^{2}_{{\rm
flat(-)}}, \label{eq3_11}
\end{eqnarray}}

\noindent where $ds^{2}_{{\rm flat(+)}}$ and $ds^{2}_{{\rm
flat(-)}}$ are respectively the form of the flat space metric for
the $z>0$ and $z<0$ domains. For $\eta >> \eta_0$  and corresponding 
$x, y, z >> a$  we may express

{\small
\begin{eqnarray}
ds^2=\frac{1}{2}\left(~{\alpha_2}+\frac{a{\alpha_1}}{r_{<}}~\right)^4~ds^{2}_{{\rm flat(+)}}
+\frac{1}{2}\left(~{\alpha_1}+\frac{a{\alpha_2}}{r_{>}}~\right)^4~ds^{2}_{{\rm
flat(-)}}~, \label{eq3_11i}
\end{eqnarray}}

\noindent and returning to Cartesian coordinates the 3-geometry can be
given in the approximate form

{\small
\begin{eqnarray}
g_{ij} \simeq \frac{1}{2}~\Big({\alpha_1}^4+{\alpha_2}^4\Big)
\Big \{1+\Big(\frac{4 a ~{\alpha_2}^3 {\alpha_1}}{{\alpha_1}^4+{\alpha_2}^4}\Big)~\frac{1}{r_{<}}+
\Big(\frac{4 a~ {\alpha_1}^3 {\alpha_2}}{{\alpha_1}^4+{\alpha_2}^4}\Big)~\frac{1}{r_{>}}\Big\}~\delta_{ij}~.
\label{eq3_12}
\end{eqnarray}}

\noindent This allows us to interpret the initial data (\ref{eq3_9})
as two interacting Schwarzschild black holes instantaneously at
rest, localized at $r_{>}(\eta_0)$ and $r_{<}(\eta_0)$, in $z>0$ and $z<0$,
with masses $M_1=2\Big\{{\alpha_1}^3 {\alpha_2}/({\alpha_1}^4+{\alpha_2}^4)\Big\}$
and $M_2=2\Big\{{\alpha_2}^3 {\alpha_1}/({\alpha_1}^4+{\alpha_2}^4)\Big\}$
in units of $a$, respectively. It is worth noticing that the
radial Schwarzschild isotropic type coordinates $r_{>}(\eta_0)$ and $r_{>}(\eta_0)$
are functions of the bispherical coordinate $\theta$, and determine the
Euclidean distance of points of the spheres $\eta=\eta_0$ to the origin.
The minimal Euclidean distance between the two spheres is given by
$L=(r_{>}+r_{<})$ evaluated along the $z$-axis, resulting in $L=2a~\exp(-\eta_0)$.

From the above construction we can now extract initial data for the
RT dynamics, which has its initial value problem on null cones.
We note that the geometry of the two spheres located at
$\eta=\eta_0$ (cf. Fig. 1) contains all the information on the
initial data for vacuum field equations through the conformal
function (\ref{eq3_9}) calculated at $\eta=0$. Based on the initial
data formulation on characteristic surfaces proposed by
D'Inverno and Stachel\cite{stachel}~\cite{vickers},
in which the degrees of freedom of the vacuum gravitational field are
contained in the conformal structure of 2-spheres embedded in a
3-spacelike surface, we are then led to adopt the conformal
structure given by

{\small
\begin{eqnarray}
K(\eta_0,\theta)=\Big( \frac{\alpha_1}{\sqrt{\cosh \eta_0 - \cos
\theta \sinh \eta_0}}+\frac{\alpha_2}{\sqrt{\cosh \eta_0 + \cos
\theta \sinh \eta_0}} \Big)^2
\label{eq3_14}
\end{eqnarray}}

\noindent as initial data for two interacting Schwarzschild black
holes to be extended along null bicharacteristics and propagated
along a timelike congruence of the spacetime. A restricted spacetime
of two interacting Schwarzschild black holes may then be constructed
locally as the product of the two 2-sphere geometry times a timelike
plane $(u,{\tilde{r}})$ generated by a null vector
$\partial/\partial {\tilde{r}}$ and a timelike vector
$\partial/\partial u$ with geometry $d
\sigma^2=\alpha^2(u,{\tilde{r}},\theta)~du^2 + 2~du~d{\tilde{r}}$.
The four geometry of the product space is then taken as

{\small
\begin{eqnarray}
ds^2= \alpha^2(u,{\tilde{r}},\theta)~du^2 + 2~du~d{\tilde{r}} -
{\tilde{r}}^2~K^2(\eta_0,\theta,u)(d \theta^2+\sin^2 \theta d
\varphi^2). \label{eq3_15}
\end{eqnarray}}

\noindent Eq. (\ref{eq3_15}) is the Robinson-Trautman metric, the
dynamics of which (ruled by Einstein's vacuum field equations)
corresponds to propagating the initial data
$K(\eta_0,\theta,u=0)=K(\eta_0,\theta)$ (cf. (\ref{eq3_14})) forward
in time from the characteristic initial surface $u=u_0$.

This characteristic initial data problem presents an ambiguity in
the interpretation of the parameter $\eta_0$. On one hand it may be
interpreted as the instantaneous boost parameter of the black holes
along the $z$ axis; on the other hand it may work as defining a
distance between the two black holes. A reason for this is that the
family of 2-spheres parametrized with $\eta_0$ (via the conformal
factor (\ref{eq3_14})) are propagated along the timelike vector
$\partial/\partial u$ that is boosted relative to the timelike
vector $\partial/\partial t$. For the simple cases of $\alpha_1=0$
or $\alpha_2=0$ this boost transformation is generated by the
conformal factor (\ref{eq3_14}) itself\cite{sachs}. We therefore are
led to interpret (\ref{eq3_14}) as representing two initially
boosted black holes, with opposite velocities $\rm v = \tanh \eta_0$
along the $z$ axis and with an initial Euclidean distance parameter $L/a =
2~\exp(-\eta_0)$.

\section{Numerical results: emission of gravitational waves and mass loss}

We now evolve the initial data (\ref{eq3_14}) via the RT equation
from $u=0$. This equation is integrated numerically using the
Galerkin method (see Refs. \cite{oliv1},\cite{oliv2} for details) in
which the approximate solution of the RT equation has the following form

{\small
\begin{equation} K(u,\theta) = A_0\,\exp\left(\frac{1}{2}
\sum_{k=0}^N\,b_k(u) P(\cos \theta)\right) \label{eq22}
\end{equation}}

\noindent where $A_0$ is an arbitrary constant, $b_k(u)$ are unknown
modal coefficients and $P_k(\cos \theta)$ are Legendre polynomials.
According to the Galerkin method a set of $N+1$ ordinary
differential equations determines the evolution of the modal
coefficients, and therefore the function $K(u,\theta)$. The initial
conditions $b_k(0)$, $k=0,1,..,N$ must correspond to the initial
data (\ref{eq3_14}) and are evaluated from

{\small
\begin{equation}
b_{k}=\frac{\left<2~\ln~(K(x) A_0^{-1}),P_{k}(\cos \theta)
\right>}{\left<P_{k}(\cos \theta),P_{k}(\cos \theta)\right>},
\label{eq23}
\end{equation}}

\noindent where the brackets define the orthogonal product in the
projection space of Legendre polynomial normalized according to
$\left<P_{k}(\cos \theta),P_{j}(\cos
\theta)\right>=\int_{0}^\pi\,P_{k}(\cos \theta),P_{j}(\cos \theta)
\sin \theta d \theta = 2\delta_{kj}/(2k+1)$. The effect of
increasing the truncation order $N$ is illustrated in Fig. 2 with
the distribution of the absolute error between the exact and
approximate initial data, respectively given by (\ref{eq3_14}) and
$K_{\rm{approx}} = \exp\left(\frac{1}{2} \sum_{k=0}^N\,b_k(0) P(\cos
\theta)\right)$.

\begin{figure}[h]
\begin{center}
\hspace{-0.1cm}
\rotatebox{270}
{\includegraphics[width=6.5cm,height=8cm]{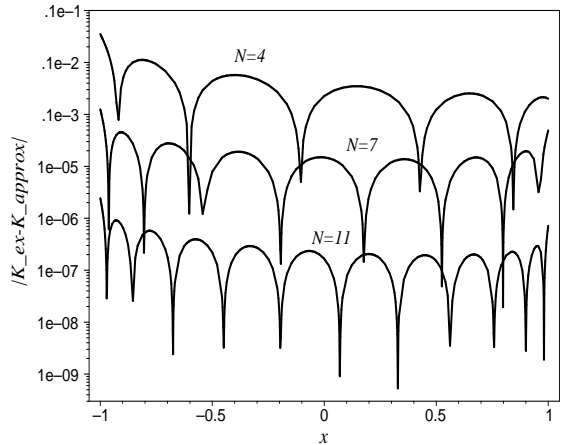}}
\caption{Plot of the absolute error between the approximate and
exact initial data for $\alpha_2=1.0$, $\alpha_1=0.2$ and
$\gamma=0.6$. The curves correspond to truncations orders
$N=4,7,11$. Here we have replaced the angular coordinated $\theta$
by $x=\cos \theta$.} \label{figB}
\end{center}
\end{figure}

We have done exhaustive numerical experiments considering truncation
order $N=13$ and taking into account distinct values for $\alpha_1$,
$\alpha_2$ and $\eta_0$. Basically, the numerical outcomes can be
summarized as:

\begin{itemize}
\item $\alpha_1 \neq \alpha_2$ (two unequal mass black holes): the final configuration is a boosted black hole with a smaller
velocity, or $\eta_{0~\rm{final}} < \eta_0$;
\\
\item $\alpha_1 = \alpha_2$ (two equal mass black holes): the final configuration is a black hole at rest, or $\eta_{0~\rm{final}}=0$.
\end{itemize}

\noindent The final configuration $K_{\rm{final}}$ can be
reconstruct as

{\small
\begin{eqnarray}
\label{eq17} K_{\rm{final}}(\theta)=\exp
\left(\frac{1}{2}\sum^N_{k=0}\,b_{k}(u_{\rm{f}}) P_{k}(\cos \theta),
\right) \label{eq24}
\end{eqnarray}}

\noindent where at the final time of integration $u = u_{\rm{f}}$,
the modal coefficients $b_{k}(u_{\rm{f}}) \approx $ constant (at
least up to $10^{-10}$).

{\small
\begin{eqnarray}
\label{eqL} K_{\rm{final}}(\theta) \simeq
\frac{K_{0\,\rm{final}}}{\cosh \eta_{0\,\rm{final}}+ \cos \theta~
\sinh \eta_{0\,\rm{final}}}.
\label{eq25}
\end{eqnarray}
}

\noindent It is worth noting that for all $K_{\rm{final}}$ it
is always possible to find values for $K_{0\,\rm{final}}$ and
$\eta_{0~\rm{final}}$ such as the inequality

{\small \begin{equation} \left| K_{\rm{final}}(\theta) -
K_{0\,\rm{final}}/(\cosh \eta_{0\,\rm{final}}+ \cos \theta~ \sinh
\eta_{0\,\rm{final}}) \right| \leq 10^{-7}
\label{eqerror}
\end{equation}
}

\noindent holds for all $0 \leq \theta \leq \pi$. Therefore the
final configuration of the system is a boosted black hole; the
final velocity and rest mass are determined and given respectively
by $v_{\rm{final}}=\tanh(\eta_{0\,\rm{final}})$ and $m_{\rm{final}}=m_0
K_{0\,\rm{final}}^3$.

One of the most interesting aspects of the dynamics of the RT
spacetimes is the emission of gravitational waves and consequently
the mass loss of the initial configuration. Let us explore this
feature considering the situation in which $\alpha_2$ and $\eta_0$ are
fixed and $\alpha_1$ is the free parameter. Therefore at $u=0$ there are two
boosted black holes directed towards each other with the same
velocity and separated by a fixed distance. One of them has fixed
mass while the mass of the second varies by changing $\alpha_1$. In
the numerical experiments we have considered the range for which $1
\ll \alpha_2/\alpha_1 \approx 1$. In particular for
$\alpha_2/\alpha_1 \ll 1$ the initial data can be recast as

{\small
\begin{eqnarray}
K(\theta) \simeq \frac{\alpha_2^2}{\cosh \eta_0 + \cos \theta \sinh
\eta_0}+\frac{2\alpha_1\alpha_2}{(\cosh^2 \eta_0 - \cos^2 \theta
\sinh^2 \eta_0)^{1/2}}, \label{eq26}
\end{eqnarray}}

\noindent meaning that it can be viewed as a boosted black hole with
total mass-energy $E_2=m_0 \alpha_2^6 \cosh \eta_0$ perturbed by a
smaller black hole whose total mass-energy is $E_1=m_0 \alpha_1^6
\cosh \eta_0$ directed towards the first black hole. Noticed that
since $E_1 \ll E_2$ its contribution to the total mass-energy
associated to the initial data can be neglected.

According to Eq. (\ref{eqBondi}) the total mass-energy of the system
evaluated at $u=0$ is given by $E_0 = 1/2
m_0\,\int_{0}^{\pi}\,K^3(\theta) \sin \theta\,d \theta$, so that
the efficiency of the process of emission of gravitational radiation
can defined by\cite{eardley}

{\small
\begin{equation}
\Delta = \frac{E_0-M_{BH}}{E_0}, \label{eq27}
\end{equation}}

\noindent where $M_{BH}$ is the total mass-energy of the resulting final
black hole. In the present context part of the kinetic energy and
the interacting energy of both black holes is radiated away, and
another part is absorbed by the first black hole increasing its rest
mass to the amount $m_0 K_{0\,\rm{final}}^3$. The RT equation is integrated
taking into account the initial data (\ref{eq3_14}) with
$\eta_0=0.3$, $\alpha_2=1$ and varying $\alpha_1$ inside the range
$[10^{-3},1.3]$. For each value of $\alpha_1$ the final outcome is
determined and identified as a boosted black hole with smaller
velocity, $v_{\rm{final}}=\tanh \eta_{0~\rm{final}} < \tanh \eta_0$
and total mass-energy such that $M_{BH}>E_1+E_2$ as expected from
the area theorem of black holes\cite{hawking1}. In Fig. 3 the result
of the numerical experiments for which $10^{-3} \leq \alpha_1 \leq
0.3$ is displayed by the plot of the efficiency $\Delta$ as function
of $M_{BH}$, along with a continuous line accounting for an
analytical function that fits quite well the numerically generated
points. Then, as we have shown previously\cite{oliv3} the function
$\Delta=\Delta(M_{BH})$ inspired from non-extensive
statistics\cite{tsallis} is given by

\begin{equation}
\Delta = \Delta_{\rm{max}}\,(1-y^\gamma)^{1/1-q}, \label{eq28}
\end{equation}

\noindent where $y=E_2/M_{BH}$ is the ratio between the initial
mass-energy of the first black hole (fixed parameter $\alpha_2$) and
the mass-energy of the resulting black hole, $\Delta_{\rm max}$ is
the maximum efficiency of the process attained in the limit $M_{BH}
\gg E_2$; $\gamma$ and $q$ are the free parameters that characterize
the non-extensive\cite{tsallis} relation. The best fit shown in Fig.
3 demands $\gamma \simeq 0.525$, $q \simeq 0.502$ and
$\Delta_{\rm{max}} \simeq 0.000977$, which is a quite small
efficiency even for a considerably high initial boost of $\tanh
(0.3) \approx 0.3$ or about 90,000 km/s. As a matter of fact we
expect that the maximum efficiency will depend on the initial boost
of both black holes. In any case the maximum efficiency is in
accordance with the value about 0.07\% found by Smarr and
Eppley\cite{eardley} for the collision of two equal and non-rotating
black holes in a more general spacetime. Although the non-extensive
relation (\ref{eq28}) is in excellent agreement with the numerical
results in the range of $\alpha_1$ under consideration that covers
the interval $M_{BH}^{\rm{min}} \leq M_{BH} \lesssim
5.0~M_{BH}^{\rm{min}}$, where $M_{BH}^{\rm{min}}$ is the minimum
value of the mass of the final black hole, it fails to describe the
entire set of points. The numerical value of the maximum efficiency,
$0.000598$, obtained when $\alpha_1=\alpha_2$ is distinct from the
one predicted by Eq. (\ref{eq28}). Also for 
$\alpha_1 \gtrsim \alpha_2$ the efficiency seems to decrease,
as would be expected from the symmetry in the interchange of
$\alpha_1$ and $\alpha_2$ in the initial data considered.

\begin{figure}
\begin{center}
\rotatebox{270}{\includegraphics*[height=8cm,width=6.5cm]{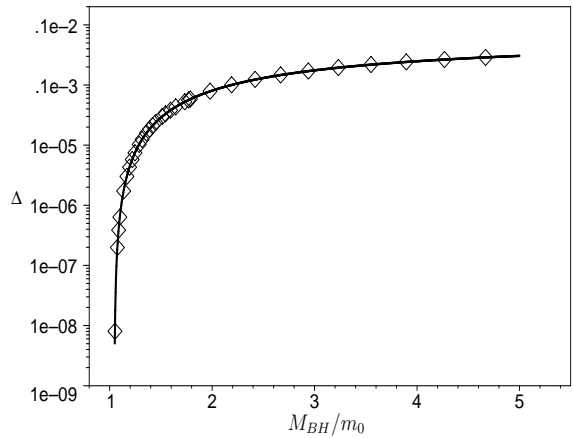}}
\caption{Plot of the ratio $M_{BH}/m_0$ versus the efficiency
$\Delta$ in the process of gravitational wave emission due to the
collision of two black holes for $10^{-3} \leq \alpha_1 \leq 0.3$
and $\eta_0=0.3$. The continuous line is the best fit of the points
by the function (\ref{eq28}) with $q \simeq 1/2$.}
\end{center}
\end{figure}

\vspace{-0.5cm}

In spite of the limitation of the non-extensive relation over entire
range of numerical points, we cannot underestimate the fact expressed by the
fitting of Fig. 3. Let us briefly discuss some implications of relation
(\ref{eq28}), starting from the domain of $\Delta/\Delta_{\rm{max}}
\ll 1$, or relatively very small efficiency.
In this situation it is clear that $E_2 \sim M_{BH}$ and therefore
$y \sim 1$, so that Eq. (\ref{eq28}) can be written as the following
scaling relation

\begin{equation}
M_{BH} \approx
E_2\,\left(1+\frac{1}{\gamma}\left(\frac{\Delta}{\Delta_{\rm{max}}}\right)^{1-q}\right).
\label{eq30}
\end{equation}

\noindent Based on the linearized theory of gravitational wave
emission we can accordingly demonstrate that the parameter $q$ must
be 1/2, the same value obtained from our numerical experiments in
the full nonlinear regime of RT dynamics. Actually, in the limit
$\Delta/\Delta_{\rm{max}} \ll 1$ we might consider the initial mass
given by $E_0=E_2+\delta E$, where $\delta E/E_2 \equiv \epsilon \ll
1$. From the quadrupole formula\cite{mtw} the mass loss can be
written as $\partial \delta E/\partial t \sim {\cal{O}}(\epsilon^2)$
which is negative definite. This means that the perturbation falls
into the hole such that at the end its mass becomes
$M_{BH}=E_2+\delta E_{\rm{abs}}$, where $\delta E_{\rm{abs}}$ is the
total amount of mass absorbed. The other fraction of the
perturbation $\delta E_{\rm{rad}}$ is radiated away, and as a
consequence of the quadrupole formula it follows that $\delta
E_{\rm{rad}}/E_2 \sim {\cal{O}}(\epsilon^2)$. Taking into account
these relations we can express the efficiency as $\Delta = (\delta E
- \delta E_{\rm{abs}})/E_2 \sim {\cal{O}}(\epsilon^2)$, and $1/y-1
\sim {\cal{O}}(\epsilon)$, which leads to

\begin{equation}
\frac{1}{y} - 1 \propto \Delta^{1/2}\;\; \Longrightarrow \;\; M_{BH}
= E_2(1+{\rm{const.}} \Delta^{1/2}).\label{eq31}
\end{equation}

\noindent Comparing Eqs. (\ref{eq30}) and (\ref{eq31}) we find
$q=1/2$. It must be emphasized that the above derivation is general
and therefore not restricted to the realm of Robinson-Trautman
spacetimes.

\begin{figure}[htb]
\begin{center}
\rotatebox{270}{\includegraphics*[scale=0.33]{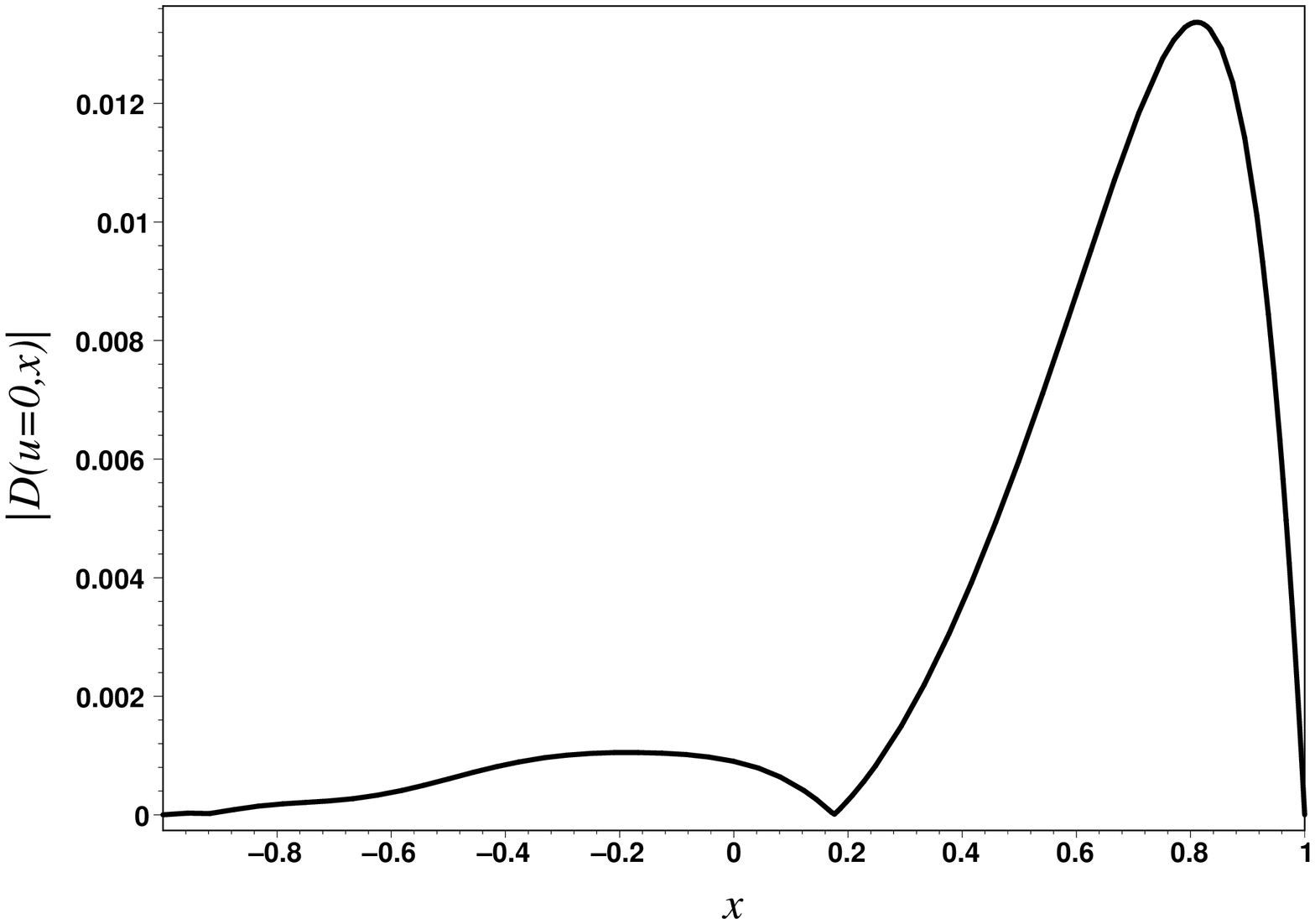}}
\rotatebox{270}{\includegraphics*[scale=0.33]{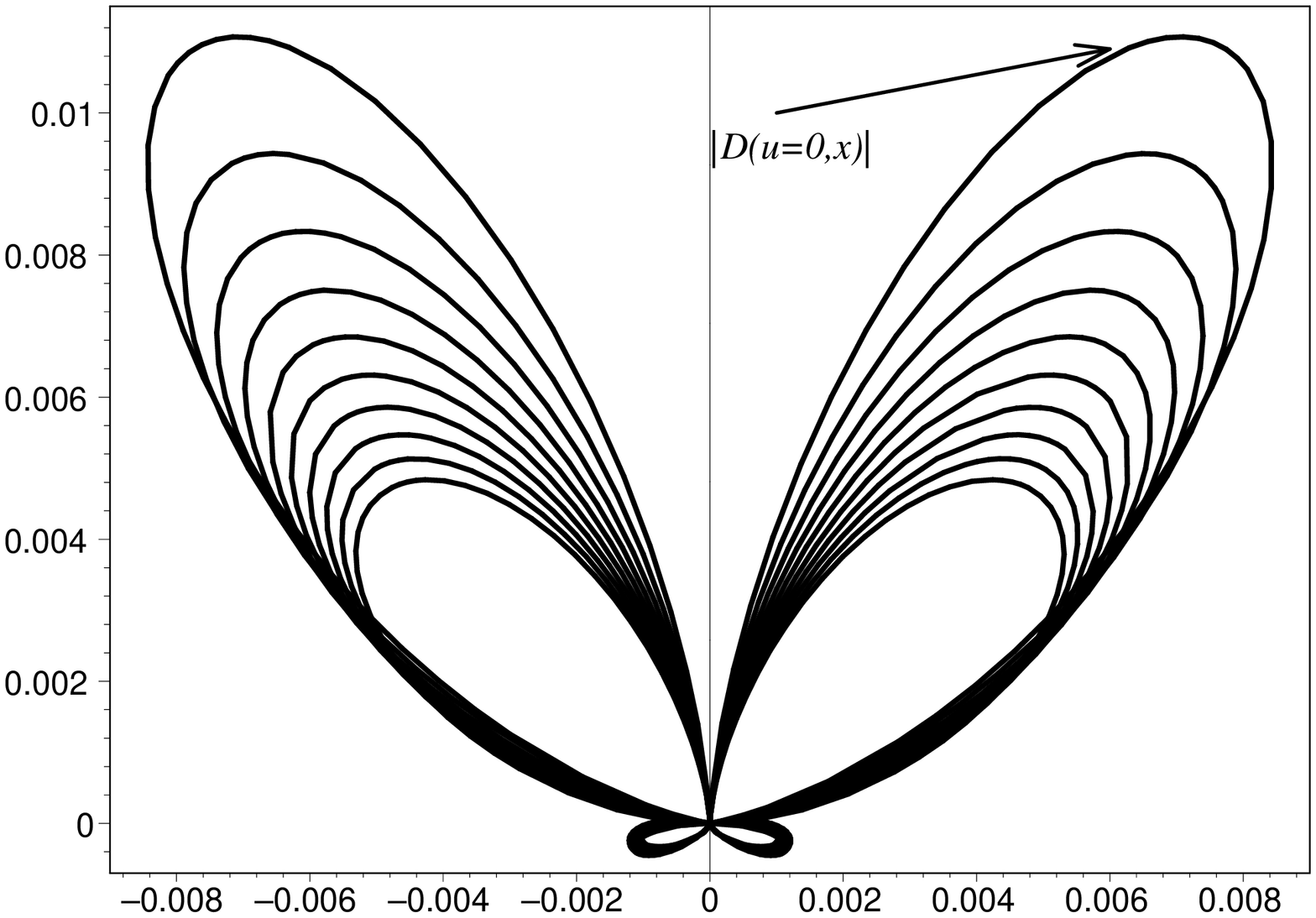}}
\caption{Initial amplitude of gravitational waves $|D(u=0,x)|$ for
which $\alpha_1=0.3$, $\alpha_2=1$ and $\eta_0=0.6$. The second
figure shows a sequence of polar plots of $|D(u,x)|$ at early
times. Notice that the angular pattern is typical of bremsstrahlung
analogous to the electromagnetic bremsstrahlung of a decelerated
charge along its direction of motion, where the lobes open due to
the deceleration. For later times the angular pattern evolves
to a typical quadrupole form.}
\end{center}
\end{figure}

The second aspect we will discuss is the wave zone angular pattern
of the gravitational waves emitted as described by the function
$D(u,\theta)$ (cf. Eq.(\ref{eq6})). Its form is valid for any value
of $r$ sufficiently large since $1/r$ is a multiplicative factor to
$D(u,\theta)$ in the wave zone curvature component. As we have
mentioned the case $\alpha_1=\alpha_2$ does not produce a boosted
black hole due to the symmetry of the initial data. In order to
illustrate the type of angular pattern of emitted waves we set
$\eta_0=0.6$ corresponding a considerably high velocity and two
black holes with initial masses of about the same order, or
$\alpha_2=1,\alpha_1=0.3$. In Fig. 4(a) we show the initial pattern
of these two black holes represented by the plot of $|D(u=0,x)|$
versus $x=\cos \theta$, that exhibits a dominant emission in the
northern hemisphere, with a maximum amplitude at $x \simeq 0.81$.
We note that the emission cone of maximum amplitude opens up due
to the decrease in the deceleration of the system; also as it opens up
the maximum amplitude decreases and the pattern evolves to a quadrupole
form for later times. The final state is that of a boosted Schwarzschild
black hole with $\eta_{0~\rm{final}} \approx 0.355$.

\section{Conclusions and Final Discussions}

In this paper we have studied the head-on collision of two
non-spinning black holes in the realm of Robinson-Trautman
spacetimes. Brill-Lindquist type initial data were
constructed that represent instantaneously two Schwarzschild black holes
directed towards each other with the same velocity.
The RT equation was integrated numerically
using a combination of Galerkin and collocation methods\cite{oliv2}
for black holes with distinct initial masses. In
general the resulting configuration is a boosted black hole as
described by Eq. (\ref{eq25}) characterized by a smaller velocity than both initial velocities, while the final mass is greater than the sum of the initial masses, or $M_{BH} > E_1 + E_2$. In this process a fraction of the total initial mass is extracted by gravitational waves, whose amount has been evaluated
for distinct values of the parameters.

Besides the derivation of the initial data the paper presents two
relevant aspects of the dynamics in the full nonlinear regime.
The first is a relation between the efficiency
$\Delta$ (cf. Eq. \ref{eq28}) of the gravitational wave extraction
and the mass of the final black hole. The remarkable feature is that
the numerical results expressed by the points $(M_{BH},\Delta)$
could be fitted by an analytical function inspired from the
non-extensive statistics as we have done considering other initial
data, with entropic index $q \simeq 1/2$. The question of whether
or not such a relation is an artifact
of Robinson-Trautman spacetimes is crucial for guiding us to further
investigation on this issue. Nonetheless we have guaranteed that at
least for very small efficiency the power-law (\ref{eq30}) derived
from the quadrupole formula is general (not restricted to Robinson-Trautman
spacetimes) and appears as a well-defined
limit of the non-extensive relation (\ref{eq28}) with $q=1/2$. Another important
aspect we have displayed is the bremsstrahlung pattern of
gravitational radiation in the wave zone, a consequence of the
deceleration of both black holes as they coalesce.

As compared to previous works (Refs.
\cite{aninos}~\cite{pp}~\cite{gleiser}~
\cite{aninos1}~\cite{aninos2}), most of which deals with the head-on
collision of two Schwarzschild black holes, our approach differs
basically in that we have adopted the characteristic surface initial
data formalism. The latter indeed has several advantages for the
description of the gravitational radiation and also for the
construction of marching algorithms\cite{winicour}. Therefore, we
consider that the problem of collision of two black holes in the
realm of characteristics is worth studying. In this direction,
an accurate code based on the Galerkin method with collocation was
constructed to integrate the field equations. The code is highly
stable for long time runs in the full nonlinear regime so that we
are able to reach numerically (up to $10^{-7}$, cf. Eq.
(\ref{eqerror})) the final configuration of the system, when the
gravitational emission ceases. The final configuration is a remnant
Schwarzschild boosted black hole, with smaller velocity parameter
($\eta_{0\,\rm{final}}<\eta_0$), and definite rest mass
$m_{\rm{final}}=m_0 K_{0\,\rm{final}}^3$ and total mass-energy
$M_{BH}=\Big( m_0
K_{0\,\rm{final}}^3~\rm{cosh}\eta_{0\,\rm{final}}\Big)$ (cf. Eq.
(\ref{eqerror})). We were therefore able to evaluate the efficiency
of gravitational radiation emission in extracting mass-energy of the
source in the full nonlinear regime as a function of the final black
hole mass and the initial mass-energy of the system. This procedure
is in the line of, and consistent with the general estimates made by
Eardley\cite{eardley}.
Our approach is alternative to the one made in the above mentioned
references on head-on black holes collisions, which evaluated the
asymptotic energy flux carried off by the gravitational wave (or
equivalently the efficiency of the gravitational wave extraction of mass from the
source) through the use of the extracted gauge invariant Zerilli
function\cite{aninos2}. The comparison of the results in the two
procedures is not straightforward due to the fact that the one
parameter Misner data and our three-parameter BL-type data are
somewhat distinct, and numerical evaluations of the efficiency were
made with variation of distinct parameters (for instance initial
separation or mass-energy of the final configuration). We should
mention that a few numerical tests made with fixed $\alpha_1=1.0$
and $\alpha_2=0.1$ and varying the boost parameter $\eta_0$ in our
approach showed a saturation tendency of the efficiency for
increasing $\eta_0$, which is in accordance qualitatively with the
second order perturbation results of Ref. \cite{nicasio}. We expect
to examine these issues in the future.

Finally, we may cite further steps in our future investigations: (i)
extension of the current work by constructing and evolving initial
data that represent a non-central collision of two black holes; (ii)
the construction of characteristic initial data that may accommodate
spinning black holes; (iii) a more general and detailed study of the
amount of mass extracted by gravitational waves and (iv) the
generality of the non-extensive relation (\ref{eq28}) connecting the
efficiency and the mass of the final black hole.

The authors acknowledge the financial support of CNPq/MCT-Brasil.
We also thank an anonymous referee for criticisms and suggestions
that allowed to improve substantially the paper.

\end{document}